\documentstyle[preprint,aps]{revtex}

\def\fpj{\hspace{-.7cm}}
\def\thalf{{\textstyle{\frac{1}{2}}}}
\def\oneth{{\textstyle{\frac{1}{3}}}}
\def\twoth{{\textstyle{\frac{2}{3}}}}
\def\tquar{{\textstyle{\frac{1}{4}}}}
\def\fiveth{{\textstyle{\frac{5}{3}}}}

\def\Km{$K^-$}
\def\nn{\nonumber}

\begin{document}
\draft
\preprint{SUNY-NTG-95-15; NUC-MINN-95-13-T}
\title{ Strangeness in  Hadronic Stellar Matter }
\author {R. Knorren and M. Prakash}
\address{
Physics Department, SUNY at Stony Brook, Stony Brook, NY 11794--3800 \\ }
\author { P. J. Ellis }
\address{ School of Physics and Astronomy, University of Minnesota,
Minneapolis, MN 55455 \\ }

\date{\today}
\maketitle
\begin{abstract}
We examine the presence of strangeness-bearing components, hyperons and kaons,
in dense neutron star matter.   Calculations are performed using  relativistic
mean field models, in which  both the baryon-baryon and kaon-baryon
interactions
are mediated by meson exchange.  Results of kaon condensation are found to be
qualitatively  similar to previous work with chiral models, if compatibility
of the kaon optical potentials is required. The presence of strangeness, be it
in  the form of
hyperons or kaons,  implies a reduction in the maximum mass and a relatively
large number of protons, sufficient to allow rapid cooling to take place.  The
need to improve upon the poorly-known couplings of the strange particles, which
determine the composition and structure of neutron stars, is stressed.  We also
discuss generic problems with effective masses in mean field  theories.
\end{abstract}
\vspace*{0.2in}

\section{INTRODUCTION}

The physical state and internal constitution of neutron stars chiefly depends
on the nature of strong interactions.  Although the composition and the
equation of state (EOS) of neutron star matter are not yet known with
certainty, QCD based effective Lagrangians have opened up intriguing
possibilities.  Among these is the possible existence of matter with
strangeness
to baryon ratio, $|S|/B$, of order unity.  Strangeness may occur in the form of
fermions, notably the $\Lambda$ and $\Sigma^-$ hyperons, or as a Bose
condensate, such as a $K^-$ meson condensate, or in the form of strange quarks
in a mixed phase of hadrons and quarks. {\em All these alternatives involve
negatively charged matter}, which if present in dense matter, results in
important consequences for neutron stars (see, for example, Ref.~\cite{pcl}).
For example, the appearance of strangeness-bearing components results in
protoneutron (newly born) stars  having larger maximum masses than catalyzed
(older, neutrino-free) neutron stars,
a reversal from ordinary nucleons-only matter. This permits the existence of
metastable protoneutron stars that could collapse to black holes during their
deleptonization~\cite{bb}.  In older stars, the presence of such components
also implies rapid cooling of the star's interior via the direct Urca
processes~\cite{rapid}.  Interpretation of the surface temperatures of neutron
stars in conjunction with different possibilities for the star's core cooling
is currently a topic of much interest~\cite{cool1}.

\vspace*{0.2in}

Our objective here is to investigate kaon condensation in dense neutron star
matter allowing for the explicit presence of hyperons.   The fact that hyperons
can significantly influence neutron star structure has been  emphasized by
Glendenning~\cite{glen} and Ellis et al.~\cite{eko}. With respect to kaons, the
suggestion of Kaplan and Nelson \cite{kapnel} that, above some critical
density, the ground state of baryonic matter might contain a Bose-Einstein
condensate of negatively charged kaons has generated a flurry of activity
examining the effective chiral Lagrangian approach and  exploring the
astrophysical consequences, e.g. [8--11]. In chiral $SU(3)_L\times SU(3)_R$ the
baryons are directly coupled to the kaons. This leads to a strong attraction
between $K^-$ mesons and baryons which increases with density and lowers the
energy of the zero-momentum state.  A condensate forms when this energy becomes
equal to the kaon chemical potential, $\mu$.

\vspace*{0.2in}

In cold catalyzed (neutrino-free) dense neutron star matter containing only
nucleons,  $\mu$ is related to the electron (or muon) and nucleon chemical
potentials by
\begin{eqnarray}
\mu = \mu_n-\mu_p =\mu_e = \mu_\mu\;,
\end{eqnarray}
due to chemical equilibrium  in the  reactions
\begin{eqnarray}
n\leftrightarrow p+e^-+\bar{\nu}_e\,, \qquad
e^- \rightarrow \mu^- + \bar{\nu}_\mu + \nu_e \qquad {\rm and} \qquad
n\leftrightarrow p+K^- \,.
\end{eqnarray}
Typically, the critical density for condensation is $\sim (3-4)n_0$ in
nucleons-only matter (where $n_0$ denotes equilibrium nuclear matter density),
although it is model and parameter dependent. It is usually the case that a
density of this order is less than the central density in a neutron star,  so a
$K^-$ condensate is expected to be present in the core region.

\vspace*{0.2in}

However, many calculations of dense matter \cite{pcl,glen,eko} indicate that
hyperons, starting with the  $\Sigma^-$ and $\Lambda$, begin to appear
at densities $\sim (2-3)n_0$.
The requirement of chemical equilibrium in the weak processes yields
\begin{eqnarray}
&&\mu_{\Lambda} = \mu_{\Sigma^0} = \mu_{\Xi^0} = \mu_n \nn\\
&&\mu_{\Sigma^-} = \mu_{\Xi^-} = \mu_n+\mu_e \nn\\
&&\mu_p = \mu_{\Sigma^+} = \mu_n - \mu_e \;.  \label{murel}
\end{eqnarray}
The above relations show that, in equilibrium, there exist only two independent
chemical potentials, $\mu_n$ and $\mu_e$, reflecting the conservation of baryon
number and electric charge.
The remaining condition is that of overall charge neutrality, namely
\begin{equation}
\sum_B q_Bn_B-n_K-n_e-n_{\mu}=0\;,
\end{equation}
where $q_B$ is the charge and $n_B$ is the number density of  baryon species
$B$.  With increasing density the concentration of negatively charged
hyperons rises so that fewer
electrons are required to maintain charge neutrality.  Consequently, the
rate of increase of the electron density $n_e$ and the chemical potential
$\mu_e$  slows, and in many cases it begins to drop.
Since $\mu_e=\mu$ governs the onset of kaon condensation,
the question of whether condensation occurs in the presence of hyperons in
stellar matter is raised.  In a previous  work\cite{ekp}, we addressed this
issue on the basis of the Kaplan-Nelson Lagrangian for the kaon-baryon
interactions and a Walecka-type relativistic field theoretical  approach for
the baryon-baryon interactions.  Using various models for the latter, it was
found that (i) the condensate threshold is sensitive to the behavior of the
scalar density; the more rapidly it increases with baryon density, the lower is
the threshold density for condensation, (ii) the presence of hyperons,
particularly the $\Sigma^-$, shifts the threshold for $K^-$ condensation to a
higher density, and (iii) in the mean field approach, with hyperons, the
condensate amplitude grows sufficiently rapidly that the nucleon effective mass
vanishes at a finite baryon density, perhaps signalling
strangeness-driven chiral restoration at high baryon density.

\vspace*{0.2in}

These findings have also raised further questions~\cite{note1}.  Among these
are issues related with (i) whether it is consistent to use the Walecka type
Lagrangian for the baryon-baryon interactions and the  chiral  Kaplan-Nelson
Lagrangian for  the kaon-baryon interactions, and  (ii) whether or not
qualitatively similar results would be obtained in more traditional approaches
to kaon-baryon interactions.  Here we address these issues by utilizing a
traditional  meson-exchange picture which can be used to generate the
kaon-baryon interactions \cite{jul} as well as the nucleon-nucleon interaction
\cite{mach}. In this case the kaons interact directly with the meson fields,
which we take here to be the $\sigma$, $\omega$ and $\rho$, and these in turn
interact with the baryons.  This meson exchange picture meshes more naturally
with the Walecka approach, which is usually used in the baryon sector. We note
that the earlier discussion of kaon condensation  using a similar approach by
Schaffner et. al. \cite{schaff}  was confined to nuclear matter,  where
$\mu=\mu_n-\mu_p=0$, whereas in neutron star matter in which weak interactions
are in equilibrium \cite{note2},  $\mu$ typically increases with density,
at least up to the density where other
hadronic negative charges appear in matter,
reaching values on the order of 200 MeV.

\vspace*{0.2in}

In Sec. II, we give the formalism and discuss the results  obtained in the
traditional meson-exchange model.   A comparison of the meson-exchange approach
with previous work employing  the Kaplan-Nelson Lagrangian is made in Sec. III
to identify   the common threads in the formalism  and similarities in the
results; see also the recent discussions of Brown and Rho \cite{chiopt} for
the case where hyperons are absent.  In Sec. IV, we offer a critique of the
model; in particular we discuss the sensitivity to  the poorly-known hyperon
couplings and difficulties that can arise with the  effective masses when
hyperons are present.  Our conclusions  are presented in Sec. V.

\section{MESON-EXCHANGE MODEL}

\subsection{Theory}

The total hadron Lagrangian is written as the sum of the baryon and the
kaon Lagrangians, ${\cal L}_H = {\cal L}_B+{\cal L}_K$.
In the baryon sector, we employ a  relativistic field theory model of
the Walecka type~\cite{sew}. We consider all charge states of the baryon octet
$B=n,p,\Lambda,\Sigma^+,\Sigma^-,\Sigma^0,\Xi^-$ and $\Xi^0$ (we shall use
the symbol $N$ for a nucleon).   Explicitly,
\begin{eqnarray}
{\cal L}_B &=& \!\sum_B\bar B\left(i\gamma^{\mu}\partial_{\mu}-g_{\omega B}
\gamma^{\mu}\omega_{\mu}-g_{\rho B}\gamma^{\mu}{\bf b}_{\mu}\cdot
{\bf t} -\!M_B+g_{\sigma B}\sigma\right)\!B \nonumber \\
&-& \tquar F_{\mu\nu}F^{\mu\nu}\!+\thalf m^2_{\omega}\omega_{\mu}\omega^{\mu}
+ \frac {\zeta}{4!}g_{\omega N}^4 (\omega_{\mu}\omega^{\mu})^2
\nonumber\\
&-&
\tquar {\bf B}_{\mu\nu}\cdot{\bf B}^{\mu\nu}+\thalf m^2_{\rho}{\bf b}_{\mu}
\cdot{\bf b}^{\mu}
\nonumber\\
&+& \thalf\partial_{\mu}\sigma\partial^{\mu}\sigma
-\thalf m^2_{\sigma}\sigma^2 - \oneth bM(g_{\sigma N}\sigma)^3
-\tquar c(g_{\sigma N}\sigma)^4\!.\label{hyp1}
\end{eqnarray}
Here $M_B$ is the vacuum baryon mass, the $\rho$-meson field is
denoted by ${\bf b}_{\mu}$,
the quantity ${\bf t}$ denotes the isospin operator which acts on the baryons,
and the field strength tensors  for the vector mesons are given by the usual
expressions:--
$F_{\mu\nu}=\partial_{\mu}\omega_{\nu}-\partial_{\nu}\omega_{\mu}$,
${\bf B}_{\mu\nu}=\partial_{\mu}{\bf b}_{\nu}-\partial_{\nu}{\bf b}_{\mu}$.
The nucleon mass, $M=939$ MeV, is included in the penultimate term to render
$b$ dimensionless.
In the Lagrangian we have included ``non-linear'' $\sigma^3$ and $\sigma^4$
terms so  that a reasonable compression modulus can be achieved for equilibrium
nuclear matter in the mean field approximation.  In some cases, we will also
explore the influence of a ``non-linear" vector interaction of the form
$(\omega_{\mu}\omega^{\mu})^2$ since this helps to
achieve a satisfactory description of the properties of finite nuclei in the
mean field approximation~\cite{Bod}.  More general couplings between the scalar
and vector fields as well as isovector non-linear couplings will be examined
in a later work.

\vspace*{0.2in}

For the kaon sector, we take a Lagrangian which contains the usual
kinetic energy and mass terms, along with the meson interactions,
\begin{eqnarray}
{\cal L}_K& =&\partial_{\mu}K^+\partial^{\mu}K^-
-(m_K^2-g_{\sigma K}m_K\sigma)K^+K^-\nonumber\\
&&\hspace{2cm}+i\left[g_{\omega K}\omega^{\mu}+g_{\rho K}b^{\mu}\right]
(K^+\partial_{\mu}K^--K^-\partial_{\mu}K^+)\;.
\label{kaonlag}
\end{eqnarray}
Here $b^{\mu}$ denotes the $\rho^0$ field and $m_K$ is the vacuum kaon mass
(which is present in the fourth term so that $g_{\sigma K}$ is dimensionless).
The scalar interaction term can be combined with the kaon mass into an
effective kaon mass defined by
\begin{equation}
	{m_K^*}^2 = m_K^2 - g_{\sigma K}m_K \sigma\;.
\end{equation}

We shall treat the kaons in the mean field approximation,
writing \cite{tpl} the time dependence
of the fields $K^{\pm}=\frac{1}{\sqrt{2}}f\theta e^{\pm i\mu t}$; thus,
$\theta$ gives the condensate amplitude.
For the baryons, we shall consider calculations at the mean field level.
We need to calculate the potential, $\Omega$, of the
grand canonical ensemble at zero temperature. It is straightforward to
obtain
\begin{eqnarray}
\frac{\Omega}{V}& =&
\thalf (f\theta)^2[{m_K^*}^2 - 2\mu(g_{\omega K}\omega_0+
g_{\rho K}b_0) - \mu^2]
+\thalf m^2_{\sigma}\sigma^2+\oneth bM(g_{\sigma N}\sigma)^3\nonumber\\
&&\fpj+\tquar c(g_{\sigma N}\sigma)^4-\thalf m_{\omega}^2\omega_0^2
-\frac{\zeta}{4!}(g_{\omega N}\omega_0)^4-\thalf
m_{\rho}^2b_0^2 +\sum_B \frac{1}{\pi^2}\int\limits_0^{k_{FB}}dk\,k^2
(E_B^*-\nu_B) \;.
\label{hyp2}
\end{eqnarray}
Here $V$ is the volume, $E_B^*=\sqrt{k^2+M^{*2}_B}$, and the baryon effective
masses are $M^*_B=M_B-g_{\sigma B}\sigma$.
The chemical potentials $\mu_B$ are given in terms of the effective chemical
potentials, $\nu_B$, by
\begin{equation}
\mu_B=\nu_B+g_{\omega B}\omega_0+g_{\rho B}t_{3B}b_0\;,\label{hyp3}
\end{equation}
where $t_{3B}$ is the $z$-component of the isospin of the baryon. The
relation to the Fermi momentum $k_{FB}$ is provided by
$\nu_B=\sqrt{k_{FB}^2+M_B^{*2}}$.

\vspace*{0.2in}

The thermodynamic quantities can be obtained from the grand potential in Eq.
(\ref{hyp2}) in the standard way; thus the baryon number density
$n_B= k_{FB}^3/(3\pi^2) $, while for kaons
\begin{equation}
n_K=(f\theta)^2(\mu+g_{\omega K}\omega_0+g_{\rho K}b_0)\;.
\end{equation}
The pressure $P=-\Omega/V$ and the energy
density $\varepsilon=-P+\sum_B\mu_Bn_B+\mu n_K$.
The meson fields are obtained by extremizing $\Omega$, giving
\begin{eqnarray}
m_{\omega}^2\omega_0 &=& -\frac{\zeta}{6}g_{\omega N}^4\omega_0^3
+\sum_Bg_{\omega B} n_B -(f\theta)^2\mu g_{\omega K}
\nonumber\\
m_{\rho}^2b_0 &=& \sum_Bg_{\rho B}t_{3B}n_B -(f\theta)^2\mu g_{\rho K}
\nonumber\\
m_{\sigma}^2\sigma &=&  -bMg_{\sigma N}^3\sigma^2
-cg_{\sigma N}^4\sigma^3+\sum_B g_{\sigma B}n^s_B
+\thalf(f\theta)^2g_{\sigma K}m_K\;.
\label{hyp5}
\end{eqnarray}
Here $n^s_B$ denotes the baryon scalar density
\begin{equation}
n^s_B=\frac{1}{\pi^2}\int\limits_0^{k_{FB}} dk\,k^2\frac{M^*_B}{E_B^*} \;.
\end{equation}
Notice that the condensate contributes directly to the equations of motion
(\ref{hyp5}), whereas in chiral models the contribution appears in the
effective chemical potentials and effective masses. Further
discussion of the two models is given in Sec. III below.

\vspace*{0.2in}

The condensate amplitude, $\theta$, is also found by extremizing $\Omega$.
This yields the solutions $\theta=0$ (no condensate), or, if a condensate
exists, the equation\cite{note3}
\begin{equation}
\mu^2+2\mu(g_{\omega K}\omega_0+g_{\rho K}b_0)-{m_K^*}^2=0\;.
\label{thresh1}
\end{equation}
The roots of this equation are the energies of the zero-momentum $K^-$ and
$K^+$ states,
\begin{equation}
\omega^\pm = \sqrt{(g_{\omega K}\omega_0+g_{\rho K}b_0)^2 + {m_K^*}^2}
	\pm (g_{\omega K}\omega_0+g_{\rho K}b_0) \;, \label{kenerg}
\end{equation}
so Eq. (\ref{thresh1}) amounts to setting the chemical
potential equal to the energy of the lowest ($K^-$) state.

\vspace*{0.2in}

Eq. (\ref{thresh1}) can be used to simplify the expressions for pressure and
energy density:
\begin{eqnarray}
P &=& -\thalf m^2_{\sigma}\sigma^2 - \oneth bM(g_{\sigma N}\sigma)^3
	- \tquar c(g_{\sigma N}\sigma)^4
	+ \thalf m_{\omega}^2\omega_0^2
        +\frac{\zeta}{4!}(g_{\omega N}\omega_0)^4
        + \thalf m_{\rho}^2b_0^2\nonumber\\
	&& + \sum_B \frac{1}{3\pi^2}\int\limits_0^{k_{FB}}dk\,
        \frac{k^4}{E_B^*} \label{pres}\\
\varepsilon &=& (f\theta)^2{m_K^*}^2
	+\thalf m^2_{\sigma}\sigma^2+\oneth bM(g_{\sigma N}\sigma)^3
	+\tquar c(g_{\sigma N}\sigma)^4
	+\thalf m_{\omega}^2\omega_0^2 +\frac{\zeta}{8}(g_{\omega N}\omega_0)^4
	+\thalf m_{\rho}^2b_0^2\nonumber\\
	&& +\sum_B \frac{1}{\pi^2}\int\limits_0^{k_{FB}}dk\,k^2
	E_B^* \;. \label{eps}
\end{eqnarray}
Note that by virtue of Eq.~(\ref{thresh1}),
the first term in the expression for the thermodynamical potential
Eq. (\ref{hyp2}) vanishes, and so the pressure due to the kaons is contained
entirely in the meson fields via their field equations (\ref{hyp5}).

\vspace*{0.2in}

To complete the thermodynamics, leptonic contributions to the total energy
density and pressure, which are given adequately by the standard  free gas
expressions, must be added to Eq.~(\ref{pres}) and Eq.~(\ref{eps}).

\subsection{Coupling Constants}

In the effective Lagrangian approach adopted here, knowledge of
three distinct sets of coupling constants is required for numerical
computations.  These are the nucleon, hyperon and kaon couplings associated
with the exchange of $\sigma$, $\omega$ and $\rho$ mesons.  In what follows, we
consider each of these in turn.

\subsubsection{Nucleon couplings}

The nucleon-meson coupling constants are determined by adjusting them to
reproduce properties of equilibrium nuclear matter. These are the saturation
density and binding energy, the symmetry energy coefficient, the compression
modulus and the Dirac effective mass at saturation.  There is a  considerable
range of uncertainty in two of the empirical values  that are to be fitted, the
compression modulus and the  Dirac effective mass, and correspondingly we
consider different sets  of coupling constants to cover this range.  The
constants determined in  this way are given in Table 1. The sets marked H are
from Heide \cite{Heide} and those labelled GM  are from Glendenning and
Moszkowski \cite{glenmos}. We also consider one  set labelled B91 with
the non-linear $\omega$ coupling~\cite{Bod}  for which the parameter
$\zeta=0.02364$. This has the advantage that one can achieve a value of
$M^*_N/M_N\sim0.6$, as favored by nuclei (in particular, spin-orbit
splittings and the charge density distributions), and a reasonable  compression
modulus with a positive value of the coefficient $c$ so that  the scalar
potential is bounded from below. We also list here for future use set HS81
taken from Ref. \cite{hs}. In addition, Table 1 gives the scalar and vector
fields at equilibrium in nuclear matter, $S = g_{\sigma N}\sigma$ and  $V =
g_{\omega N}\omega_0$, which are relevant for the calculation of the  kaon
optical potential.

\subsubsection{Hyperon couplings}

Following Glendenning and Moszkowski \cite{glenmos}, we constrain the
coupling constants of the $\Lambda$ hyperon by requiring that the correct
binding energy be obtained for the lowest $\Lambda$ level in nuclear matter
at saturation. Defining $x_{\sigma\Lambda}=g_{\sigma\Lambda/}g_{\sigma N}$,
with
analogous definitions for the $\omega$ and $\rho$ couplings, this gives
\begin{equation}
-28=x_{\omega\Lambda} g_{\omega N}\omega_0-x_{\sigma\Lambda}
g_{\sigma N}\sigma\;,
\label{hyp6}
\end{equation}
in units of MeV. We adopt the value $x_{\sigma\Lambda}=0.6$, as suggested in
Ref. \cite{glenmos} on the basis of fits to hypernuclear levels and neutron
star properties;  $x_{\omega\Lambda}$ is then determined. We also choose
$x_{\rho\Lambda}= x_{\sigma\Lambda}$, since the alternative,
$x_{\rho\Lambda}=x_{\omega\Lambda}$, gives  very similar results. Our choices
are listed in Table 2. In a recent analysis of $\Sigma^-$ atoms,  Mare\v{s} et
al.~\cite{sigmacou} find reasonable fits with $x_{\omega\Sigma}=\twoth$ and 1,
and  $x_{\sigma\Sigma}=0.54$ and 0.77, the larger values yielding  a slightly
better fit to the data.  To begin with, we use the values in Table 2, which
are close to the
smaller values of Mare\v{s} et al.,  for  all the hyperons and comment upon
different couplings for the $\Sigma$ and $\Xi$ later in Sec. IV.

\subsubsection{Kaon couplings}

In order to investigate the effect of a kaon condensate  on the equation of
state in high-density baryonic matter,  the kaon-meson coupling constants have
to be specified. Empirically-known quantities can be
used to  determine these constants, but it is important to keep in mind that
laboratory experiments give  information only about the kaon-nucleon
interaction in free space or  in nuclear matter (matter with a proton fraction
$x=\thalf$ at an equlibrium density of $n_0=0.153~{\rm fm}^{-3}$). On the other
hand, the physical setting in this work is matter in the
dense interiors of neutron  stars, i.e. infinite matter containing baryons
(nucleons and possibly  hyperons) and leptons in $\beta$-equilibrium, that has
a different composition  and spans a wide range in densities (up to central
densities $\sim8n_0$). As a consequence,  kaon-meson couplings as determined
from experiments might not be appropriate to describe the kaon-nucleon
interaction in neutron star matter, and the particular choices of coupling
constants should be regarded as parameters that have a range of uncertainty.

\vspace*{0.2in}

One possibility of experimentally determining the strength of the  kaon-nucleon
interaction is the analysis of phase shift data.
An analysis of $KN$ scattering data using a meson-exchange model  \cite{jul}
was used to
determine couplings of nucleons and kaons to $\sigma,  ~\omega, ~{\rm and}
\ \rho$ mesons. This yielded
\begin{eqnarray}
G^\sigma_{KN} &=& \frac{g_{\sigma N} g_{\sigma K}}{m_\sigma^2}
= 2.444~{\rm fm}^{2}\nonumber\\
G^\omega_{KN} &=& \frac{g_{\omega N} g_{\omega K}}{m_\omega^2}
= 4.981~{\rm fm}^{2}\nonumber\\
G^\rho_{KN} &=& \frac{g_{\rho N} g_{\rho K}}{m_\rho^2} = 1.301~{\rm fm}^{2}\;,
\label{kcoup}
\end{eqnarray}
with our Lagrangian conventions. Since only the ratio $g/m$ enters the
formalism, it is not necessary to specify the masses, and
the kaon ratios are listed in Table 3.

\vspace*{0.2in}

With these couplings and the field strengths in nuclear matter at saturation,
we can determine the value of the optical potential felt by a single kaon in
infinite nuclear matter for the present model.
Lagrange's equation for an $s$-wave \Km\ with a time dependence
$ K^- = k^-({\bf x})~e^{-iEt} $, where $ E = \sqrt{p^2 + m_K^2}$ is the
asymptotic energy, is obtained \cite{eric} from
Eq. (\ref{kaonlag}) as
\begin{eqnarray}
[\nabla^2 + E^2 - m_K^2]~k^-({\bf x}) &=& [-2(g_{\omega K}\omega_0 +
g_{\rho K}b_0)E - g_{\sigma K}m_k \sigma]~k^-({\bf x}) \nn\\
 &=& 2~m_K~U_{opt}^K~k^-({\bf x})\;.
\end{eqnarray}
In nuclear matter, $b_0 = 0$, so for a kaon with zero momentum ($E=m_K$) the
optical potential is
\begin{equation}
 U_{opt}^K \equiv S^K_{opt} + V^K_{opt}  =  - \thalf g_{\sigma K} \sigma
-g_{\omega K} \omega_0\;.
\label{uoptmeson}
\end{equation}
The value of $U^K_{opt}$ for the different models are contained in
Table 4.
Friedman et al. have recently reanalyzed the kaonic atom data \cite{batty}
examining a more general parameterization of $U_{opt}^K$ in nuclear matter
than the standard $t_{eff}\rho$ approximation. They were able to obtain a
better fit with a kaon optical potential whose real part had a depth of
$- 200 \pm 20 ~{\rm MeV}$.
The coupling constants adjusted to the parameters obtained from
phase shift measurements lead to a value of the kaon optical  potential close
to this value. Note that the ratio of the  $K\omega$ coupling to the
$N\omega$ coupling, $x_{\omega K} =  g_{\omega K}/g_{\omega N}$, is close to
one, whereas  $x_{\sigma K}$ and $x_{\rho K}$ are close to the value of
$\oneth$ which is suggested by naive quark counting.

\subsection{Results}

Since the kaon coupling constants are uncertain,  some
orientation is gained by plotting the threshold density for condensation
as a function of $g_{\sigma K}/m_{\sigma}$ and $g_{\omega K}/m_{\omega}$
(the threshold is less
sensitive to $ x_{\rho K}$, which we fix to be $\oneth$). Fig. 1 shows
contours of the
critical density ratios $n_{crit}/n_0$ for kaon condensation in matter
containing nucleons, while Fig. 2 shows the corresponding results for matter
containing nucleons and hyperons. Comparison of these figures shows that the
threshold density is higher when hyperons are present, and this is particularly
marked for smaller values of the couplings. For orientation, the values in
Table 3 are roughly $g_{\sigma K}/m_{\sigma}\sim0.7$ and
$g_{\omega K}/m_{\omega}\sim2$, so while  condensation will occur in
the range $u=2-3$ when hyperons are absent, their presence may increase $u$
quite substantially depending on the precise value of
$g_{\sigma K}/m_{\sigma}$. When we choose parameters as in Sec. III below,
$g_{\omega K}/m_{\omega}\sim0.7$, yielding a higher threshold, particularly
when hyperons are allowed.

\vspace*{0.2in}

In the remainder of this section, we adopt the kaon couplings of Table 3
and first discuss the nucleons-only case, followed by the
case where hyperons are also allowed.

\subsubsection{Matter containing nucleons and leptons}

In Fig. 3 we display our results for matter containing nucleons and leptons for
a representative case, parameter set GM2 with kaon couplings from Table 3. The
particle fractions, $Y_i=n_i/n$, are shown in panel 1 of Fig. 3. The proton
fraction becomes much closer to the neutron fraction once kaons are present,
and for  high $u$ they are essentially equal. It can be seen that the threshold
condition for kaon condensation, Eq. (\ref{thresh1}), is fulfilled at a density
of $2.6 n_0$. (The dashed lines in Fig. 3 show the behavior if kaons are
excluded.) In panel 2 of this figure, the energies of the zero-momentum kaon
states, $\omega^{\pm}$, are plotted as a function of the ratio of baryon
density to equilibrium nuclear matter density, i.e. $u=\sum_B n_B/n_0\equiv
n/n_0$. We see that the $\omega^-$ energy drops with increasing density and
meets the  chemical potential $\mu$ at threshold. The effective kaon mass
$m_K^*$ does not vary greatly; so, referring to Eq. (\ref{kenerg}), the
density-dependent  contributions, dominated by the term containing the $\omega$
meson, are critical in obtaining condensation. Panel 3 (right scale) shows that
the condensate amplitude, $\theta$, rises rapidly at threshold and then slowly
approaches a maximum value of $\sim40^{\circ}$; this is smaller than in  chiral
models. The effect of the  condensate on the $\omega$ and $\sigma$ fields
(panel 2) follows from the  field equations (\ref{hyp5}), whereas the behavior
of the $\rho$ field is dominated by the changes in the neutron--proton ratio.
The proton charge is balanced by  an approximately equal number of $K^-$ mesons
beyond threshold, since the lepton contributions rapidly become negligible.
Thus the magnitude of the strangeness/baryon, $|S|/B$, in panel 3 is
$\sim\thalf$ once kaons condense. Panel 4 of Fig. 3 shows that the total
pressure and energy density are reduced when kaons are present.

\vspace*{0.2in}

Finally, in the upper part of Table 5, the gross properties of neutron stars
are given for the various equations of state. The critical density ratios lie
in a narrow range, $u_{crit}\sim2.5-3$, so that a significant region of the
star will contain kaons. This  softens the equation of state
causing a reduction in the maximum mass by 4--10\%. The precise value
depends on the magnitude of the $\omega$ repulsion at high density, which
is governed by the coupling constants of Table 1. This also affects the
changes in the central density; usually this is increased by kaons, but
in the B91 case there is a small reduction.

\subsubsection {Matter containing nucleons, hyperons and leptons}

We now consider the case where hyperons are allowed to be present in addition
to nucleons. The results displayed in Fig. 4 can be compared with those of Fig.
3 where hyperons were excluded.  The particle fractions are shown in panel 1.
The first strange particle to  appear is the $\Sigma^-$, since the
somewhat higher mass of the $\Sigma^-$ is compensated by the electron chemical
potential in the equilibrium condition of the $\Sigma^-$ (see
Eq.~(\ref{murel})).  Since the $\Sigma^-$ carries a negative charge, it causes
the  lepton fractions to drop. This means that the chemical potential $\mu$ is
reduced, requiring a smaller value of $\omega^-$ for kaon condensation which
results in a higher threshold density. This is evident from
panel 2.   Also shown in panel 2 are the changes in the meson fields arising
from kaon condensation, and these are much smaller than in the absence of
hyperons. This arises partly from the reduction in the condensate amplitude
(the maximum value of $\theta$ is $\sim20^{\circ}$, see panel 3) and partly
from  changes in the baryon fractions (see panel 1).

\vspace*{0.2in}

Immediately above threshold  the kaon  fraction rises dramatically to reach a
maximum of $0.1-0.2$ per baryon. Since the kaons carry negative charge, charge
neutrality for the system leads to a small drop in the $\Sigma^-$ fraction, and
the lepton concentrations become even smaller. By contrast, the fraction of the
neutral $\Lambda$ is  little influenced by kaon condensation. In fact this is
the largest fraction  at large values of $u$, with roughly comparable amounts
of $n,p,\Sigma^-$ and a relatively small kaon presence. Thus, the hyperons
dominate the strangeness/baryon, $|S|/B$, of $\sim0.6$ at the highest density
considered (panel 3).

\vspace*{0.2in}

The pressure and energy density are displayed in panel 4 of Fig. 4.   A
comparison with the corresponding panel of Fig. 3 shows the effect of hyperons,
which, for a given baryon density,  leads to significantly smaller
pressures and larger energy densities. Panel 4 of Fig. 4 also shows
that when hyperons are present the effects of a kaon
condensate  are rather small.  The change in the energy density due to
kaons receives positive contributions from the mesons, and a large negative
contribution from the baryons. The lepton contribution is negligible. The net
result is, as it must be, a reduction in the energy density; but it is at
most only 0.2\%. Turning to the change in the pressure arising from
condensation, we first note that since we are plotting against density, rather
than chemical potential, there is no requirement as to the sign. In fact, we
see that at the lower densities the pressure is lowered (softer EOS),
whereas at the higher densities it increases (stiffer EOS). This arises
from competition between the negative $\sigma$ and $\rho$ meson contributions
and the  $\omega$ meson contribution, which is positive.

\vspace*{0.2in}

The neutron star properties for matter containing nucleons, hyperons and
leptons are given in the lower part of Table 5. Here, a single asterisk
indicates  that the neutron effective mass becomes zero before reaching the
expected central density; kaons have not condensed prior to this point.
Further discussion is given in Sec. IV below. For the other cases, and
excluding kaons for the
moment, we see that the softening effect of hyperons causes a reduction of
$\sim0.5M_{\odot}$ and a corresponding increase in the central density, as is
well known \cite{glen,eko}. If we include kaons in the calculation,
condensation takes place within the star
only for models GM2 and GM3, and it does so at a
higher density than when hyperons are absent. The reduction in the maximum
neutron star mass due to the presence of kaons amounts to only about $0.01
M_{\odot}$. The change in the  central density is likewise small.

\vspace*{0.2in}

Thus, in this model, the influence of the hyperons is decisive.
In nucleons-only
matter, the pressure is significantly decreased by a kaon condensate, which
lowers the  maximum mass.  An even larger reduction in the maximum mass
occurs  when hyperons are present in matter.  The additional presence of
condensed kaons
in hyperonic matter induces relatively small changes in the EOS  so that
there is little influence of the condensate on the gross stellar properties.

\section{COMPARISON WITH CHIRAL MODELS}

In previous work \cite{ekp}, we employed the same baryon Lagrangian
Eq.~(\ref{hyp1}), but took the kaon kinetic energy and mass terms as well as
the
kaon-nucleon and kaon-hyperon interactions from the Kaplan--Nelson chiral
Lagrangian \cite{kapnel}. We would like to compare this with the meson-exchange
approach discussed in the preceding section.  We choose to  establish
parameters that are in some sense compatible in the two cases via the optical
potential. For the chiral model, the kaon Lagrangian in {\it nuclear} matter
($n_n = n_p$) takes the form
\begin{equation}
{\cal L}_K = \partial_\mu K^+ \partial^\mu K^- - m_K^2 K^+K^- +
\frac{3i}{8f_\pi^2}
n(K^+\partial_0K^- - K^-\partial_0K^+) +
\frac{\Sigma^{KN}}{f_\pi^2}n^sK^+K^-\;,
\end{equation}
where the pion decay constant $f_\pi=93$ MeV, $n=n_n+n_p$, $n^s=n_n^s+n_p^s$
and
the kaon-nucleon sigma term $ \Sigma^{KN} = -(a_1/2 + a_2 + 2a_3)m_s$, in terms
of the standard parameters of the chiral Lagrangian \cite{kapnel}.
It is straightforward \cite{chiopt} to show that the optical potential is
\begin{equation}
	U_{opt}^{chK} \equiv S^{chK}_{opt} + V^{chK}_{opt}
= - \frac{\Sigma^{KN} n^s}{2m_K f_\pi^2} - \frac{3n}{8f_\pi^2}\;.
\label{uoptchiral}
\end{equation}
The kaon-nucleon sigma term requires the parameters $a_1m_s$, $a_2m_s$
and $a_3m_s$. The first two of these can be established from the
hyperon-nucleon mass differences, but the third is
related to the strangeness content of the proton, which is unknown. Taking the
reasonable range of 0, 10 and 20\% strangeness for the proton yields
$\Sigma^{KN}=167$, 344 and 520 MeV, respectively.
This gives $S^{chK}_{opt} = -22, -45 {\rm~and~} -69$ MeV for the
different choices. The value $V^{chK}_{opt} = -51$ MeV is given uniquely by
the saturation density. Thus, the values of the optical potential are
$U^{chK}_{opt} = ~-73, ~-96 {\rm~and~} -120$ MeV.
Due to the fact that the vector part of the potential is only about
$\oneth$ of the value in Sec. II, the total optical potential is only about
half of the value favored by Friedman et al. \cite{batty}, although it is
comparable to the value obtained in their $t_{eff}\rho$ approximation.
It must, however, be borne in mind that there are uncertainties in their
analysis and also in simply expropriating the real part of a complex
potential as we have done. Further, our main interest here is in a comparison
of the chiral model with the meson-exchange model.

\vspace*{0.2in}

We choose the coupling constants of the meson-exchange model such that the
scalar and vector parts of the optical potential  as given in Eq.
(\ref{uoptmeson}) are equal to the corresponding chiral values in Eq.
(\ref{uoptchiral}). This does not determine the kaon-rho coupling for which we
take  $ x_{\rho K} = 1/3 $. The values of the coupling constants thus
determined are given in  Table 6. Comparison with Table 3 shows that the
$\omega$ coupling is substantially reduced, whereas the $\sigma$ coupling is
increased for the larger values of $\Sigma^{KN}$.

\vspace*{0.2in}

Before proceeding, a comparison of  the expressions for the critical densities
obtained in the meson-exchange and chiral models is useful.  Both can be
written in the form
\begin{equation}
\mu^2+2\mu\alpha-m_K^{*2}=0\;.
\end{equation}
For simplicity, we
restrict ourselves to the case in which only the $\Lambda$ and $\Sigma^-$
hyperons are considered in addition to nucleons. In the chiral model,
$\alpha$ and $m_K^{*2}$ are
\begin{eqnarray}
\alpha&=&\frac{2n_p+n_n-n_{\Sigma^-}}{2f_{\pi}^2}\nonumber\\
m_K^{*2}&=&m_K^2+\left[2a_1n^s_p+(2a_2+4a_3)(n^s_p+n^s_n+n^s_{\Sigma^-})
+\left(\fiveth(a_1+a_2)+4a_3\right)n^s_{\Lambda}\right]\frac{m_s}{2f^2_{\pi}}
\;.
\end{eqnarray}
In the meson-exchange model, we have
\begin{eqnarray}
\alpha&=&\left(G^{\omega}_{KN}-\thalf G^{\rho}_{KN}\right)n_n
+\left(G^{\omega}_{KN}+\thalf G^{\rho}_{KN}\right)n_p
+G^{\omega}_{K\Lambda}n_{\Lambda}
+\left(G^{\omega}_{K\Lambda}- G^{\rho}_{K\Lambda}\right)n_{\Sigma^-}\nonumber\\
m_K^{*2}&=& m_K^2+G^{\sigma}_{KN}m_K\left[bM(g_{\sigma N}\sigma)^2
+c(g_{\sigma N}\sigma)^3-n^s_n-n^s_p\right]
-G^{\sigma}_{K\Lambda}m_K(n^s_{\Lambda}+n^s_{\Sigma^-})\;,
\end{eqnarray}
where we have used the definitions of Eq. (\ref{kcoup}).
Comparing these expressions for the two models we see that the weightings of
the various densities are different.
In addition, the ``non-linear'' $b$ and $c$ terms do not play a role in
the chiral expressions. So already the threshold condition is different in the
two approaches, even though they use the same underlying baryonic model and
give the same optical potential in nuclear matter.
Above threshold, $\theta$ enters in different ways in the two models and this
will introduce additional differences.

\vspace*{0.2in}

The properties of neutron stars in the meson-exchange model are shown in  Table
7 with and without kaons and hyperons; here, the parameters are chosen on the
basis of the optical potentials, as discussed above, and we focus on the GM
cases. These results can be
directly compared with those of the chiral model in Table 8 (note the values
listed here differ slightly from those of Ref. \cite{ekp}, where an equilibrium
nuclear matter density of 0.16 fm$^{-3}$ was employed). When kaons are
excluded, the models are, of course, identical. In the case that hyperons
are absent, the threshold for
condensation, $u_{crit}$, is noticeably lower in the chiral model by
0.2--0.7 units of the density ratio. Nevertheless, the results for the
maximum masses and central densities in the two models are, for the most
part, similar. This would suggest that our
procedure of adjusting the couplings via the optical model is reasonable.

\vspace*{0.2in}

Turning to the case where hyperons are present, the meson-exchange model
only yields a condensate for the largest value of  $\Sigma^{KN}$. This is
the bottom row of Table 7. Only for parameters GM2 and GM3 is the critical
density less than the central density; but, even in these cases,
the kaon condensation produces only a minor modification of the stellar
properties. For the chiral case in Table 8, kaon condensation occurs for
$\Sigma^{KN}$ values of 344 and 520 MeV. For the latter, the critical
densities are much lower than for the meson exchange model.  We recall
from Fig. 2 that, for $g_{\omega K}/m_{\omega}\sim0.7$, the
critical density is very sensitive to the precise value of the $\sigma$
coupling. Thus, it is to be expected that when
hyperons are present, the question of kaon condensation is quite delicate
and depends sensitively on the parameters employed, as well as
the model chosen.  Finally, as the notation in Table 8 indicates,
the effective mass drops to zero in all the chiral cases; the problem is
more severe here than in meson-exchange models, because there is an explicit
negative contribution from the condensate to the baryon masses.
Thus, we are unable to compare the stellar properties of the two models.

\section{CRITIQUE OF THE MODELS}

In this section, we want to point out clearly that there are significant
uncertainties and difficulties associated with these models. We first  discuss
the implications of uncertainties in the hyperon coupling constants,
and then we delineate difficulties with effective masses.

\subsection{Hyperon couplings}

In the previous sections, we  assumed that the couplings of the $\Sigma$ and
$\Xi$ were equal to those of the $\Lambda$ hyperon. Here, we relax this
assumption and
explore the sensitivity to unequal couplings  of the different hyperons.
Of the many possibilities, we pick three for study. These are listed in
Table 9, in terms of the ratio to the nucleon couplings as defined in
Sec. II.B.2. For the $\Lambda$, we use the values discussed previously.
For the $\Sigma$, we use two sets of values which gave satisfactory fits to the
$\Sigma^-$ atom  data in the work of Mare\v{s} et al., \cite{sigmacou}.
This was based on a mean field description of
nuclear matter using the nucleon couplings of Horowitz and Serot~\cite{hs},
who did not include non-linear terms ($b=c=\zeta=0$). The parameters are
listed in Table 1 as HS81. Partly for consistency and partly because this
model is often used as a baseline in the literature,
we will adopt these parameters. (Qualitatively similar results are obtained
for  other values of the nucleon
couplings, which yield more realistic values of the compression modulus.)
Finally, we need the couplings of the $\Xi$. Since there is little
information, we take the couplings to be equal to those of either the
$\Lambda$ or the $\Sigma$. Note that set 1 in Table 9 is close to the set
that we have been using in the previous discussion.

\vspace*{0.2in}

In Fig. 5, the upper, center and lower panels refer to hyperon coupling sets 1,
2 and 3, respectively. The upper panel is similar to results already
discussed; note that kaons do not condense up to the maximum density displayed,
$u=4.5$. In discussing the other cases, we first mention the  seeming paradox
that increasing the coupling constants of a hyperon species delays its
appearance to a higher density. The explanation  \cite{glen,eko} is that the
threshold equation receives contributions from  the $\sigma,\ \omega$ and
$\rho$ mesons, the net result being positive due to  the $\omega$. Thus, if all
the couplings are scaled up, the positive contribution becomes larger, and the
appearance of the particle is delayed to a higher  density. With this in mind,
consider the center panel of Fig. 5 which corresponds to set 2 of Table 9. The
$\Sigma$ couplings are larger than set 1 (upper panel), so the $\Sigma^-$ no
longer appears, thus allowing the chemical potential $\mu$ to continue  rising
(cf. Figs. 3 and 4). This allows the $\Xi^-$ to appear at $u=2.2$,  essentially
substituting for the $\Sigma^-$. Of course, were we to reduce the  $\Xi$
couplings on the grounds that this hyperon contains two strange quarks, the
$\Xi^-$ would appear at an even lower density. Turning to the lower panel of
Fig. 5, we recall that this  corresponds  to set 3 of Table 9, for which both
the $\Sigma$ and $\Xi$ couplings are increased. Neither of them now appear, and
since the chemical potential, $\mu$, continues to increase with density, it
becomes favorable for kaons to condense at  $u=3.6$; the fraction $Y_{K^-}$,
however, remains rather small.

\vspace*{0.2in}

Clearly, the lesson to be drawn from this is that the thresholds for the
strange particles, hyperons and kaons, are sensitive to coupling  constants
which are poorly known. Thus, while strangeness plays a significant role in
determining the constitution and physical properties of a neutron star, the
detailed behavior cannot be tied down at the present time.

\subsection{Effective Masses}

We have several times alluded to effective masses going to zero, and we wish to
clarify the situation here; for clarity kaons will be excluded from the
initial discussion. The situation is best illustrated by reference to Fig. 6.
Here, we display for all the parameter sets we have discussed (see Table 1)
the effective mass ratio,  $M_n^*/M$, for the neutron, since this is the
first  particle to show pathological behavior.  The top panel is for
the case in which only nucleons are allowed, and  we see that there is no
pathological behavior with the effective mass going  smoothly to zero with
increasing density in all cases. Indeed, in pure  neutron matter, with
degeneracy $\gamma=2$, or in nuclear matter, with $\gamma=4$,  at high density
the effective mass has the limiting form~\cite{sew}
\begin{eqnarray}
\label{mstar}
M_n^* \rightarrow M \left[ 1 + \frac {g_{\sigma N}^2}{m_\sigma^2}
\frac {\gamma k_{Fn}^2}{4\pi^2} \right] ^{-1}\;,
\end{eqnarray}
when the non-linear couplings are neglected, $b=c=0$. By contrast,
when hyperons are allowed with the couplings of Table 2, the middle
panel of Fig. 6 shows that in most cases the neutron effective mass becomes
zero. Even for the GM2 and 3 cases, this will happen if one goes beyond the
density range plotted. The density at which $M_n^*$ becomes zero is
clearly correlated with the effective nucleon mass in equilibrium nuclear
matter. Values $\sim0.6$, as favored by nuclei \cite{hs}, cause this to
happen at $u\sim4$, while for $M^*_n/M\sim0.8$, it is postponed to $u>10$.
The B91 model behaves differently, with the neutron  mass becoming zero
at $u\sim7.5$, even though the effective mass in nuclear matter is 0.6.
This behavior of the effective mass turning negative has
been noted earlier by L\'{e}vai et al. \cite{lev} in a mixture of nucleons and
$\Delta$ baryons at finite temperature and chemical potential.

\vspace*{0.2in}

The problem of effective masses turning negative is generic to  multi-component
systems in which the constituents have dissimilar masses  and different
couplings to the $\sigma$ field.   This can be seen clearly if one considers
the $\sigma$ field equation (\ref{hyp5})  (again for simplicity, choosing
$b=c=0$).  At very high densities,  one can take the limit $k_{FB}/M_B^*\gg 1$
for all baryons present.  Defining
\begin{eqnarray}
G^\sigma_B &=& g_{\sigma B} g_{\sigma N}/m_\sigma^2\,, \qquad
x_B = g_{\sigma B}/g_{\sigma N} \qquad {\rm and} \qquad
y_B = M_B/M\;,
\end{eqnarray}
the $\sigma$ field equation can be rewritten as
\begin{equation}
\frac{g_{\sigma N}\sigma}{M} =   \frac{\sum_B G^\sigma_B y_B k_{FB}^2}
	{\sum_B G^\sigma_B x_B k_{FB}^2}
	\left(\frac{1}{1+\frac{2\pi^2}
	{ \left(\sum_B G^\sigma_B x_B k_{FB}^2\right) }}\right)\;.
\end{equation}
For sufficiently high densities, the term in the bracket will approach unity.
Since $M_n^*/M=1-g_{\sigma N}\sigma/M$, the neutron effective mass will
eventually become negative if the term in front of the bracket is greater than
unity. This is the case for the middle panel of Fig. 6, since $y_B\geq1$ and
$x_B\leq1$. (Note that instead of taking the nucleon coupling and mass as a
reference, one could have taken any one of the baryon couplings and masses.)
Thus, one of the baryon effective masses will go negative at high densities
unless all the baryon couplings are chosen to fulfill
\begin{equation}
x_B = y_B\,, \qquad {\rm or} \qquad g_{\sigma B}/g_{\sigma N} = M_B/M\,.
\end{equation}
This prescription yields the following ratios for the hyperon to
nucleon coupling constants:
\begin{eqnarray}
x_{\sigma \Lambda} = 1.190 \,, \qquad
x_{\sigma \Sigma} = 1.272 \qquad {\rm and} \qquad
x_{\sigma \Xi} = 1.405\,.
\label{xs}
\end{eqnarray}
The bottom panel of Fig. 6 shows that, with these values of $x$, the effective
neutron mass now goes to zero only at infinite density, as in  nucleons-only
matter.  Similar behavior is obtained for the other baryons.  It must be
emphasized that while the choice of couplings in Eq.~(\ref{xs}) leads to
physically sensible effective masses, it fails to  reproduce the $\Lambda$
binding in Eq.~(\ref{hyp6}), unless a  significantly larger value for
the $\omega$ coupling is used.

\vspace*{0.2in}

We would like to briefly assess the implications of Eq.~(\ref{xs})  for the
composition of neutron stars.  For present purposes, we will use the $x$
values of Eq.~(\ref{xs}) for all the meson-hyperon couplings. Results are shown
in Fig. 7 for the parameter  sets H300 and GM2; the latter can be compared with
Figs. 3 and 4. The upper panels (without kaons) show that hyperons appear, but
that the fractions $Y$ of the various species are small and tend to drop with
increasing density. The matter is dominantly neutron matter. This is reflected
in the maximum mass for the GM2 case, which is 2.04$M_{\odot}$, essentially the
same as the np case of  Table 5. If one allows for the presence of kaons, with
the parameters of  Table 3, the lower panels of Fig. 7 show they appear at
$u\sim3$ and quickly balance the number of protons, $Y_{K^-}\approx Y_p$, and
this is close to the neutron fraction, $Y_n$, at high density. Thus, the matter
is dominantly npK matter, and again this is reflected in the maximum mass for
the GM2 case, which is $1.86M_{\odot}$ (cf. Table 5). In a nutshell, with these
values of $x$, hyperons play only a very minor role.

\vspace*{0.2in}

\newcommand{\vmg}[1]{\mbox{\boldmath$#1$}}

The behavior of the effective masses, namely  that they vanish at a finite
baryon density, may indicate that the mean field model is being pushed to the
limits of its applicability.  Inclusion of quantum loop corrections and the
effects of correlations may well alter this behavior.   Vanishing effective
masses naturally arise in models with spontaneously broken chiral symmetry,
where the vacuum masses are generated by a non-zero value for a scalar field.
(The Nambu--Jona-Lasinio model, with four-Fermi interactions of the type
$(\bar\psi\psi)^2-(\bar\psi\gamma_5 \vmg {\tau}\psi)^2$,
is an example.)   These
models also suggest that at a finite density chiral symmetry is restored, i.e.
the effective  masses become zero, even for a single fermion species.  In the
Walecka type model we are considering this occurs  at infinite density for
nucleons of zero strangeness, but at a finite density when  fermions of
different strangeness enter. Whether this may be interpreted as {\em
strangeness-induced chiral restoration} in this approach depends on (i)
whether the effective mass can be viewed as an  order parameter, and (ii) also
on whether the scalar terms of the Walecka  Lagrangian can be shown to arise
from chiral Lagrangians.  While arguments  to support such an interpretation
may be adduced, further work is clearly necessary.

\section{CONCLUSIONS}

We have examined the presence of strangeness in the form of hyperons and/or
a kaon condensate in neutron star matter.  Calculations were performed in the
framework of a relativistic  mean field theory in which  baryon-baryon and
kaon-baryon interactions are generated by the exchange of $\sigma,\omega$ and
$\rho$ mesons.   Our results allow for comparisons with the results obtained
from the chiral Kaplan-Nelson model, where the kaons and  baryons interact
directly.

\vspace*{0.2in}

The qualitative results of kaon condensation in  the meson-exchange model are
similar to those of the chiral model when the magnitudes of the kaon-baryon
interactions in the two models are required to be compatible with the kaon
optical potential in nuclear matter.   Specifically, we find that  when matter
contains nucleons and leptons only, kaons condense around  $3-4$ times the
nuclear matter  saturation density, as found earlier using the chiral model.
The effects of kaon condensation include a softening of the equation of state,
which leads to a reduction in the maximum mass and to an increased proton
concentration, which implies a rapid cooling of the star's core through direct
Urca (beta decay) processes.

\vspace*{0.2in}

Within the meson-exchange model, we also investigated the presence of hyperons
in matter and its influence on kaon condensation.  Due to limited guidance
about the couplings of the strange particles, be they kaons or hyperons,  the
densities at which they appear in dense matter are uncertain.    The
importance of individual hyperon species likewise remains unclear.  For
example, with the choice of couplings made in Sec. II (suggested by the
$\Lambda$ binding in nuclei), hyperons play a dominant role, while kaons, which
appear at a higher density, are of lesser importance. However, these roles  may
be altered by other suitable choices of the coupling constants.  If, in
addition to the $\Lambda$ couplings implied by hypernuclei, the couplings
implied by $\Sigma^-$ atoms are employed, then,  depending upon the couplings
of the $\Xi$, it is possible that the $K^-$ is the first  negatively charged
hadron to appear in
matter.   This highlights the importance of further work in this area using
inputs from hypernuclear physics and advances in both theory and techniques
for calculating the energy of interacting systems.

\vspace*{0.2in}

In addition, we have pointed out that in mean field theories, effective masses
will become zero and negative, unless a particular choice is made for the
couplings of the hyperons to the $\sigma$ field;  this choice is, however,
not supported by data on hypernuclei. Even if this problem does not occur
within the range of densities considered, an instability in the basic theory
gives one cause for concern in assessing the results obtained in these and
other
calculations. Thus, there is need to devise alternative means to
treat a system of many baryons in the presence of scalar interactions.

\vspace*{0.2in}

Despite these caveats, an overall theme does emerge. Namely, that the presence
of strangeness in dense matter necessarily implies that the equation of state
is softened and the maximum mass is reduced.   As noted in Ref.~\cite{pcl},
protoneutron stars  with strangeness-bearing components  have larger maximum
masses than catalyzed neutron stars, in contrast to  the case of nucleons-only
stars.  This leads to  metastable protoneutron stars that could collapse to
black holes during their deleptonization era.  In addition, since strangeness
is accompanied by negative charge, the proton fraction in processed stars  will
be large enough for the direct Urca process to operate.  This leads to more
rapid cooling than the ``standard'' modified Urca process, which requires a
spectator nucleon. The rate of cooling can be determined from
the inferred surface temperatures of neutron stars \cite{cool1},
but further work is needed before a definitive answer can be given.

\vspace*{0.2in}

We refer the interested reader to a related paper by Schaffner and
Mishustin~\cite{sm95}, which became available as this work was completed.

\vspace*{0.2in}

\section*{Acknowledgements}
We thank Gerry Brown and Mannque Rho for much encouragement and valuable
discussions. Two of us (MP and PJE) thank the Institute for Nuclear Theory
at the University of Washington for its hospitality during part of this work.
This work was supported in part by the U. S. Department of Energy
under grant numbers DE-FG02-88ER40388 (RK and MP) and DE-FG02-87ER40328 (PJE).
\vspace*{0.2in}

% =============== References ===================================

% ===================== Tables: =================================

\newpage
\begin{quote}
TABLE 1. Coupling constants fitted to a binding energy of $-16.3$ MeV at an
equilibrium density of $n_0=0.153~{\rm fm}^{-3}$ in nuclear matter with a
compression modulus $K$ and effective mass $M^*$.  The symmetry energy
coefficient is 32.5 MeV.  The equilibrium scalar and vector fields are
also listed. Models termed `H' are from Ref.~\cite{Heide},
models termed `GM' are from Ref.~\cite{glenmos}. For the model termed
`B91' from Ref.~\cite{Bod} and the model termed `HS81' from Ref.
{}~\cite{hs}, the binding energy is $-15.75$ MeV at $n_0=0.1484~{\rm fm}^{-3}$
and the symmetry energy coefficient is 35 MeV.
\end{quote}
\begin{center}
\begin{tabular}{c|cc|ccc|cc|cc}
\hline \hline
{}	& $M^*_N/M_N$ & $K$ &
$ \displaystyle{\frac{g_{\sigma N}}{m_\sigma}}$ &
$ \displaystyle{\frac{g_{\omega N}}{m_\omega}}$ &
$ \displaystyle{\frac{g_{\rho N}}{m_\rho}}$ & $b$ & $c$ &$S$ & $V$ \\
{}	&{}	& (MeV)	& (fm) & (fm) & (fm) & & & (MeV) & (MeV)\\ \hline
H300& 0.65 & 300 & 3.655 	& 2.932	& 2.035	& 0.002319 & -0.002129
& 329 & 260\\
H200& 0.65 & 200 & 3.758	& 2.932	& 2.035	& 0.003845 & -0.005035
& " & " \\ \hline
GM1 & 0.7  & 300 & 3.434	& 2.674	& 2.100	& 0.002947 & -0.001070
& 282 & 216 \\
GM2 & 0.78 & 300 & 3.025	& 2.195	& 2.189	& 0.003478 &  0.01328
& 207 & 146 \\
GM3 & 0.78 & 240 & 3.151	& 2.195	& 2.189	& 0.008659 & -0.002421
& " & " \\ \hline
B91 & 0.6  & 250 &  4.068	& 3.392 & 2.173 & 0.0014028 & 0.0001193
& 376 & 304 \\ \hline
HS81  & 0.541  & 545 &  3.974	& 3.477 & 2.069 & 0.0 & 0.0
& 431 & 354 \\
\hline \hline
\end{tabular}
\end{center}
\vskip.5cm
\begin{quote}
TABLE 2. Ratios of hyperon-meson to nucleon-meson coupling constants.
\end{quote}
\begin{center}
\begin{tabular}{c|c|ccc} \hline \hline
 	& $M^*_N/M_N$ & $x_{\sigma H}$ & $x_{\omega H}$ & $x_{\rho H}$\\ \hline
H200,300	& 0.65 	& 0.6 & 0.652 & 0.6\\ \hline
GM1		& 0.7  	& 0.6 & 0.653 & 0.6\\
GM2,3 		& 0.78	& 0.6 & 0.659 & 0.6\\ \hline
B91 		& 0.6 	& 0.6 & 0.649 & 0.6\\ \hline
HS81 		& 0.541 & 0.6 & 0.651 & 0.6\\
\hline \hline
\end{tabular}
\end{center}

\newpage
\begin{quote}
TABLE 3.  Kaon-baryon coupling constants that reproduce the phase shift data.
\end{quote}
\begin{center}
\begin{tabular}{c|ccc}
\hline \hline
{} &
$ \displaystyle{\frac{g_{\sigma K}}{m_\sigma}}$ &
$ \displaystyle{\frac{g_{\omega K}}{m_\omega}}$ &
$ \displaystyle{\frac{g_{\rho K}}{m_\rho}}$ \\
{}      & (fm) & (fm) & (fm) \\ \hline
H300    & 0.669 & 1.699 & 0.639 \\
H200    & 0.650 & 1.699 & 0.639 \\ \hline
GM1    & 0.712 & 1.863 & 0.619 \\
GM2    & 0.808 & 2.269 & 0.594 \\
GM3    & 0.776 & 2.269 & 0.594 \\ \hline
B91 	& 0.601 & 1.468 & 0.599 \\ \hline
HS81 	& 0.615 & 1.433 & 0.629 \\
\hline \hline
\end{tabular}
\end{center}
\vskip.5cm
\begin{quote}
TABLE 4. Scalar and vector contributions and the kaon optical potential in
equilibrium nuclear matter using the kaon-baryon couplings from the phase
shifts.
\end{quote}
\begin{center}
\begin{tabular}{c|ccc}
\hline \hline
{}	& $-S^K_{opt}$   & $-V^K_{opt}$ & $-U^K_{opt}$ \\
{}	& (MeV) & (MeV) & (MeV) \\ \hline
H300  	&  30 & 150 & 180 \\
H200  	&  28 & 150 & 178 \\ \hline
GM1 	&  29 & 151 & 180 \\
GM2 	&  28 & 151 & 179 \\
GM3 	&  25 & 151 & 176 \\ \hline
B91 	& 28 & 132 & 159 \\ \hline
HS81	& 33 & 146 & 179 \\
\hline \hline
\end{tabular}
\end{center}

\newpage
\begin{quote}
TABLE 5. Gravitational mass and central density of the maximum mass
neutron stars for matter with and without kaon condensates.   The critical
density ratio for condensation is given in the middle column. The symbol np
denotes matter  containing nucleons and leptons, and npH denotes matter
containing nucleons, hyperons and leptons.
%rk:
The kaon coupling constants are taken from Table 3. The symbol * marks models
for which the nucleon effective mass drops to zero before reaching the central
stellar density. The symbol ** indicates that for this choice of
constants no condensation takes place up to the maximum density considered
($u=10$).
\end{quote}
\begin{center}
\begin{tabular}{c|c|cc|c|cc}
\hline \hline
 & & & & & & \\
 &  	& \multicolumn{2} {c|} {\raisebox{1.5ex}[2ex]{without kaons}} &
 & \multicolumn{2}{c} {\raisebox{1.5ex}[2ex]{with kaons}}
\\ \hline
 & 	& $\displaystyle{\frac{M_{max}}{M_\odot}}$ & $u_{cent}$ & $u_{crit}$ &
$\displaystyle{\frac{M_{max}}{M_\odot}}$ & $u_{cent}$\\ \hline
	& H300	& 2.529 & 5.13 & 2.44 & 2.395 & 5.56 \\
	& H200	& 2.508 & 5.32 & 2.40 & 2.378 & 5.81 \\ \cline{2-7}
 	& GM1 	& 2.346 & 5.70 & 2.49 & 2.185 & 6.46 \\
np 	& GM2 	& 2.064 & 6.58 & 2.60 & 1.854 & 8.37 \\
 	& GM3 	& 2.005 & 7.14 & 2.59 & 1.809 & 9.18 \\ \cline{2-7}
	& B91 	& 2.097 & 5.80 & 3.19 & 2.014 & 5.59 \\ \cline{2-7}
	& HS81  & 2.954 & 3.85 & 3.35 & 2.886 & 3.86 \\ \hline \hline
 	& H300 	& ---   & *    &   ** & ---  & ---  \\
 	& H200 	& ---   & *   &    ** & ---  & ---  \\ \cline{2-7}
 	& GM1 	& 1.776 & 6.53 &   **  & ---  & ---  \\
npH 	& GM2 	& 1.655 & 6.96 & 3.37 & 1.645 & 7.36 \\
 	& GM3 	& 1.544 & 7.98 & 3.20 & 1.536 & 8.46 \\ \cline{2-7}
        & B91 	& 1.463  & 6.18 & ** & --- & --- \\ \cline{2-7}
        & HS81  & ---   & *   &  **  & ---   & --- \\
\hline \hline
\end{tabular}
\end{center}

\newpage
\begin{quote}
TABLE 6. Kaon-baryon coupling constants determined from the optical
potential in chiral models. For $\rho$-meson exchange,
we take the ratio $x_{\rho K} = g_{\rho K}/g_{\rho N}$ to be $\oneth$.
\end{quote}
\begin{center}
\begin{tabular}{c|ccc|c|c}
\hline \hline
{} & {}	&
$ \displaystyle{\frac{g_{\sigma K}}{m_\sigma}}$ &  {} &
$ \displaystyle{\frac{g_{\omega K}}{m_\omega}}$ &
$ \displaystyle{\frac{g_{\rho K}}{m_\rho}}$\\
{}		&	& (fm) &	& (fm) & (fm)   \\ \hline
 $\Sigma^{KN}$ (MeV)	& 167	& 344	& 520	& 	&	\\ \hline
GM1		& 0.537 & 1.106 & 1.672	& 0.631	& 0.700	\\
GM2		& 0.650 & 1.340 & 2.025	& 0.769	& 0.730	\\
GM3		& 0.678 & 1.396 & 2.110 & 0.769 & 0.730 \\
\hline \hline
\end{tabular}
\end{center}

\newpage
\begin{quote}
TABLE 7.  Gravitational mass and central density of the maximum mass neutron
stars for matter with and without kaon condensates in the meson-exchange model.
The critical density ratio for condensation is also listed.   Symbols are: np
for nucleons-only matter, npK for nucleons-only matter with a kaon condensate;
npH and  npHK denote matter which also includes hyperons.  The kaon coupling
constants are from Table 6. The symbol ** indicates  that no condensation takes
place for this choice of coupling constants.
\end{quote}
\begin{center}
\begin{tabular}{cc|ccc|ccc|ccc}
\hline \hline
 & & & GM1 & & & GM2 & & & GM3 & \\
 & & $\displaystyle{\frac{M_{max}}{M_\odot}}$  & $u_{cent}$ & $u_{crit}$
& $\displaystyle{\frac{M_{max}}{M_\odot}}$  & $u_{cent}$ & $u_{crit}$
& $\displaystyle{\frac{M_{max}}{M_\odot}}$  & $u_{cent}$ & $u_{crit}$ \\
 &$\Sigma^{KN}$& & & & & & & & & \\ \hline\hline
np  &     & 2.346 & 5.70 & ---    & 2.064 & 6.58 & ---    & 2.005 & 7.14 & ---
\\ \hline
    & 167 & 2.339 & 5.62 & 4.74 & 2.038 & 6.22 & 4.90 & 1.950 & 6.66 & 4.75\\
npK & 344 & 2.288 & 5.56 & 3.74 & 1.956 & 5.97 & 3.95 & 1.831 & 7.82 & 3.75\\
    & 520 & 2.177 & 6.35 & 2.94 & 1.818 & 8.85 & 3.17 & 1.770 & 9.25 & 2.99
\\ \hline \hline
npH  & 	  & 1.776 & 6.53 & ---    & 1.655 & 6.96 & --- & 1.554 & 7.98 & --- \\
\hline
     & 167 & ---  & ---  &  **  &  ---  & ---  & **   &  ---  &  --- & ** \\
npHK & 344 & ---  & ---  &  **  &  ---  & ---  & **   &  ---  &  --- & ** \\
     & 520 & ---  & ---  & 7.72 & 1.646 & 6.89 & 5.10 & 1.516 & 8.35 & 3.93\\
\hline \hline
\end{tabular}
\end{center}

\newpage
\begin{quote}
TABLE 8. Same as Table 7, but for the chiral model.    The symbol * indicates
that the effective mass drops to zero below the expected central stellar
density. The critical density for condensation is marked with a ** if the
nucleon effective mass drops to zero prior to condensation.
\end{quote}
\begin{center}
\begin{tabular}{cc|ccc|ccc|ccc}
\hline \hline
 & & & GM1 & & & GM2 & & & GM3 & \\
 & & $\displaystyle{\frac{M_{max}}{M_\odot}}$  & $u_{cent}$ & $u_{crit}$
& $\displaystyle{\frac{M_{max}}{M_\odot}}$  & $u_{cent}$ & $u_{crit}$
& $\displaystyle{\frac{M_{max}}{M_\odot}}$  & $u_{cent}$ & $u_{crit}$ \\
 &$\Sigma^{KN}$& & & & & & & & & \\ \hline\hline
np  &     & 2.346 & 5.70 & ---    & 2.064 & 6.58 & ---    & 2.005 & 7.14 & ---
\\ \hline
    & 167 & 2.334 & 5.60 & 4.54 & 1.990 &  5.96 & 4.33 & 1.911 &  6.67 & 4.35\\
npK & 344 & 2.270 & 5.66 & 3.48 & 1.796 &  8.68 & 3.28 & 1.783 &  9.22 & 3.29\\
    & 520 & 2.182 & 6.45 & 2.73 & 1.769 & 10.39 & 2.60 & 1.777 & 10.30 & 2.61
\\ \hline \hline
npH  & 	  & 1.776 & 6.53 & ---    & 1.655 & 6.96 & --- & 1.554 & 7.98 & --- \\
\hline
     & 167 & --- & --- &  **  & --- &  *  & 9.39 & --- &  * & 9.90\\
npHK & 344 & --- &  * & 5.87 & --- &  *  & 4.39 & --- &  * & 4.41\\
     & 520 & --- &  * & 3.33 & --- &  *  & 2.86 & --- &  * & 2.86\\
\hline \hline
\end{tabular}
\end{center}
\newpage
\begin{quote}
TABLE 9. Ratios of hyperon-meson to nucleon-meson coupling constants,
$x_{iH}=g_{iH}/g_{iN}$ where $i=\sigma,\ \omega$ or $\rho$ and $H$ is a
hyperon species.
\end{quote}
\begin{center}
\begin{tabular}{c|ccc|ccc|ccc} \hline \hline
Case & $x_{\sigma \Lambda}$ & $x_{\omega \Lambda}$ & $x_{\rho \Lambda}$ &
$x_{\sigma \Sigma}$ & $x_{\omega \Sigma}$ & $x_{\rho \Sigma}$ &
$x_{\sigma \Xi}$ & $x_{\omega \Xi}$ & $x_{\rho \Xi}$ \\ \hline
1 & 0.60 & 0.65 & 0.60 & 0.54 & 0.67 & 0.67 & 0.60 & 0.65 & 0.60\\
2 & 0.60 & 0.65 & 0.60 & 0.77 & 1.00 & 0.67 & 0.60 & 0.65 & 0.60\\
3 & 0.60 & 0.65 & 0.60 & 0.77 & 1.00 & 0.67 & 0.77 & 1.00 & 0.67\\
\hline \hline
\end{tabular}
\end{center}

% ===================== Figure captions: ========================

\newpage
\section*{Figure captions}

\begin{quote}
Fig. 1. Contours of the critical density ratio  $u_{crit}= n_{crit}/n_0$ for
kaon condensation  in matter containing nucleons and leptons as a function of
the  kaon-meson coupling constants $g_{\omega K}/m_\omega$ and  $g_{\sigma
K}/m_\sigma$. For the kaon-rho coupling the ratio $x_{\rho K} = g_{\rho
K}/g_{\rho n}$ was taken to be $\oneth$.  Panels (1)--(3) show results for
models GM1-GM3.
\end{quote}

\begin{quote}
Fig. 2. Same as Fig.1, but in matter containing nucleons, hyperons and leptons.
In region (A), kaons are present above the critical density up to the highest
densities considered ($8n_0$ for panel (1) and $10n_0$ for panels (2) and
(3)). In region (B), kaons appear at the critical density but disappear again
at a higher density. In this region, for each choice of couplings the figure
yields two densities, the lower one corresponding to the critical density for
kaon condensation and the higher one to the highest density where kaons will
still be present. In region (C), kaons do not condense.
\end{quote}

\begin{quote}
Fig. 3. Matter containing nucleons and leptons with parameter set GM2.
Solid (dashed) lines show quantities in matter with (without) kaons,
as a function of the baryon density ratio $u=n/n_0$.
Panel (1): Particle fractions $Y_i=n_i/n$.
Panel (2): Kaon energies $\omega^\pm$ and effective mass $m_K^*$,
meson field strengths and  electron chemical potential $\mu$.
Panel (3): Kaon condensate amplitude, $\theta$ and strangeness/baryon,
$|S|/B$. Panel (4): Pressure and energy density.
\end{quote}

\begin{quote}
Fig. 4. As for Fig. 3, but for matter containing nucleons, hyperons and
leptons.
\end{quote}

\begin{quote}
Fig. 5. Particle fractions for model HS81 with different choices of
$\Sigma$ and $\Xi$ coupling constants. Panels (1), (2) and (3)
correspond to parameter sets 1, 2 and 3 of Table 9, respectively.
\end{quote}

\begin{quote}
Fig. 6. Neutron effective mass ratios, $M^*_n/M$, in nucleons-only matter
(panel 1), in matter containing hyperons with the parameters of Table 2
(panel 2) and in matter containing hyperons with the parameters of
Eq. (\ref{xs}) (panel 3). The labels on the curves indicate the
equation of state employed (see Table 1).
\end{quote}

\begin{quote}
Fig. 7. Particle fractions in the case where the hyperon couplings to all
mesons are chosen according to Eq. (\ref{xs}). The left panels show
results for the HS300 parameter set and the right panels for the GM2 set.
In the upper (lower) panels kaons are excluded from (included in) the
calculations.
\end{quote}

\end{document}